\documentclass[fleqn,usenatbib]{mnras}

\usepackage{newtxtext,newtxmath}
\usepackage{subcaption}
\usepackage{xcolor}
\newcommand{\mega}{\mbox{\sc mega}}
\newcommand{\skirt}{\mbox{\sc skirt}}

\newcommand{\eagle}{\mbox{\sc{Eagle}}}

\newcommand{\jwst}{\mbox{\it JWST}}
\newcommand{\flares}{\mbox{\sc Flares}}

\newcommand{\um}{\SI{}{\micro\meter}}
\newcommand{\mumetre}{\SI{}{\micro\meter}}
\usepackage{siunitx}
\usepackage{gensymb}

\usepackage{placeins} 
\defcitealias{10.1093/mnras/staa3360}{\flares\,I}
\defcitealias{10.1093/mnras/staa3715}{\flares\,II}
\defcitealias{Vijayan_2022}{\flares\,III}
\defcitealias{10.1093/mnras/stac1368}{\flares\,IV}

\usepackage[T1]{fontenc}

\DeclareRobustCommand{\VAN}[3]{#2}
\let\VANthebibliography\thebibliography
\def\thebibliography{\DeclareRobustCommand{\VAN}[3]{##3}\VANthebibliography}

\usepackage{graphicx}	
\usepackage{amsmath}	

\usepackage{ulem}
\usepackage{float}

\title[FLARES XXII]{First Light and Reionization Epoch Simulations (FLARES) XXII: UV-dust spatial offsets at the Epoch of Reionisation}

\author[P. Punyasheel et al.]{Paurush Punyasheel,$^{1}$\thanks{E-mail: z.paurush-punyasheel@herts.ac.uk}
Aswin P. Vijayan,$^{2}$
William J. Roper,$^{2}$
Thomas R. Greve,$^{3,4}$
Hiddo Algera,$^{5}$ \newauthor
Christopher C. Lovell,$^{6,7}$
Steven Gillman,$^{3,4}$
Bitten Gullberg,$^{3,4}$
Shihong Liao,$^{8}$
Robert M. Yates$^{1}$\newauthor
and Stephen M. Wilkins$^{2}$ \newauthor
\\
$^{1}$Centre for Astrophysics Research, Department of Physics, Astronomy and Mathematics, University of Hertfordshire, College Lane, Hatfield AL10 9AB, UK \\
$^{2}$Astronomy Centre, University of Sussex, Falmer, Brighton BN1 9QH, UK\\
$^{3}$DTU Space, Technical University of Denmark, Elektrovej 327,DK-2800 Kongens Lyngby, Denmark \\
$^{4}$ Cosmic Dawn centre (DAWN), Copenhagen, Denmark\\
$^{5}$ Institute of Astronomy and Astrophysics, Academia Sinica, 11F of Astronomy-Mathematics Building, No.1, Sec. 4, Roosevelt Rd, Taipei 106319, Taiwan, R.O.C. \\
$^{6}$ Kavli Institute for Cosmology, University of Cambridge, Madingley Road, Cambridge CB3 0HA, UK \\ 
$^{7}$ Institute of Astronomy, University of Cambridge, Madingley Road, Cambridge CB3 0HA, UK\\
$^{8}$ Key Laboratory for Computational Astrophysics, National Astronomical Observatories, Chinese Academy of Sciences, Beijing 100101, China \\
}

\date{Accepted XXX. Received YYY; in original form ZZZ}

\pubyear{\the\year{}}

\begin{document}
\label{firstpage}
\pagerange{\pageref{firstpage}--\pageref{lastpage}}
\maketitle

\begin{abstract}
Recent observations have revealed intriguing offsets between the UV and FIR emission in high redshift galaxies. In this study, we use the First Light And Reionisation Epoch Simulations (\textsc{Flares}) to compute the spatial offset of ultraviolet (UV) and far-infrared (FIR) centres for a statistical sample (6890) of massive (M$_{\star}\, \gtrsim10^{9} \,{\rm M_{\odot}}$) high redshift galaxies ($z \in [5,10]$). The galaxies are post-processed with the \textsc{skirt} radiative transfer code, to obtain the full spectral energy distribution and surface brightness profile.  We simulate \textit{James Webb Space Telescope (JWST)} Near Infrared Camera (NIRCam; rest-frame 1500 \AA , $ \approx 0.031 ''$ resolution) and ALMA rest-frame 158 \um\ ($\approx$ $0.3''$ angular resolution) observations of the galaxies and then calculate the distance between the UV-FIR centres to analyse which physical processes drive the observed UV - FIR spatial offset. We find that $\sim16.23\%$ of galaxies exhibit spatial offsets of $\geq 2.5$ kpc between their UV and FIR emission peaks. We establish that the spatial offsets do not correlate with stellar mass, UV/FIR luminosity, and size. Offsets also do not correlate with AGN feedback or with large-scale environment or merger history. Galaxies with significant offsets preferentially have bluer UV slopes ($-2.5<\beta<-1.5$), consistent with recent star formation and dust-attenuated cores displacing the observed UV centroid. They show an accelerated star formation history, forming half their $z=5$ stellar mass $\sim$0.1 Gyr earlier than galaxies without offsets. These galaxies are enriched earlier than galaxies without an offset and show enhanced stellar metallicities, indicating a transition to an outward growth at higher redshifts ($z \geq 6$). 
\end{abstract}




\section{Introduction}

Dust in any astrophysical system impacts the radiation
travelling through it by scattering and absorbing photons \citep{Salim_2020}. Accounting for this dust attenuation is critical in astronomical observations \cite[e.g.][]{2024MNRAS.528..997Q, 2025ApJ...990..114S, 2025arXiv250720190V} to derive accurate physical properties. The ultraviolet (UV) spectrum, dominated by the emission from young, bright O/B stars, is highly susceptible to attenuation, which absorbs and re-emits this energy in the infrared (IR). By jointly studying UV and IR emission, we can probe the processes driving the stellar mass growth of galaxies meticulously. With the advent of high-resolution rest-frame UV continuum imaging from \textit{JWST}, now extending such capabilities to galaxies at the Epoch of Reionisation (EoR), together with dust continuum detections from ALMA at comparable resolutions (< 1 kpc), the time is ripe for panchromatic studies of early galaxies.

Dust-obscured star formation is an important component of galaxy evolution and has been extensively studied across a wide range of redshifts \cite[$z \sim 1$--5, e.g.][]{Casey_2014,Dunlop17,Bouwens2020}. At intermediate redshifts ($z \sim 2$--3), submillimetre galaxy (SMG) samples trace the most dust-rich and intensely star-forming systems. At higher redshifts ($z \gtrsim 4$), ALMA surveys such as ALPINE \citep{2020A&A...643A...1L,2020A&A...643A...2B,2020ApJS..247...61F} and CRISTAL \citep{mitsuhashi2023almacristal}, along with programs like AS2UDS \citep{stach19,10.1093/mnras/stz2835}, have begun to probe more typical main-sequence star-forming galaxies. However, such observations are expensive, making it challenging to assemble statistically significant samples. Dust-obscured star formation remains significant at $z \approx 7$ \citep{Algera_2022}, and blind ALMA surveys have identified heavily obscured, NIRCam/optically dark galaxies \citep{2021Natur.597..489F,2025arXiv250606418S,NIRCamDark}.

With the increasing spatial resolution of ALMA and \jwst, these observations now enable measurements of the spatial extent of dust and stellar emission in high-redshift galaxies. \citet{mitsuhashi2023almacristal} measure the sizes of typical star-forming galaxies at $z=4$--6 and find that the half-light radii of dust emission at 158~\um\ are approximately twice those of the UV continuum. At $z \approx 8$, \citet{ULRIG} show that dust attenuation inferred from \jwst\ observations underestimates the total dust obscuration compared to ALMA measurements, suggesting a spatial separation between stars and dust.

Interpreting these observations, however, remains challenging \citep{2023A&A...677A..34G,2024MNRAS.530..966G}. Current observational samples are often limited in size and biased towards the brightest or most dust-obscured systems, while intrinsic galaxy structure and dust geometry cannot be directly constrained observationally. Simulations provide a complementary framework to interpret these observations. They enable the study of large, statistically representative galaxy samples, including systems that may be missed by observational selection effects \citep{2011A&A...534A..81R}, and allow direct access to intrinsic galaxy properties that are otherwise difficult to constrain. By applying observational pipelines to simulated galaxies, it is possible to directly connect intrinsic structure with observed quantities \citep{10.1093/mnras/staa137,Vijayan_2022,Popping_2021,10.1093/mnras/stab3794,2025ApJ...978L..42C,Paurush}. Cosmological hydrodynamical simulations provide a complementary approach to studying dust-obscured star formation.

Within this framework, \flares\ \citep{10.1093/mnras/staa3360,10.1093/mnras/staa3715} predicts intrinsically compact dust emission, with dust half-light radii more than five times smaller than their UV counterparts \citep{10.1093/mnras/stac1368}. However, when these same galaxies are passed through a simulated observation pipeline, the ratio of ALMA (158~\um) to FUV \jwst/NIRCam half-light sizes instead suggests dust sizes that are roughly twice as large \citep{Paurush}. Similarly, using TNG50 simulations, \citet{Popping_2021} report intrinsic IR-to-UV size ratios of less than unity. These theoretical predictions, which stand in tension with observations, highlight the importance of disentangling intrinsic galaxy structure from observational biases when interpreting UV and IR continuum sizes together. Doing so requires studies that apply realistic observational pipelines to theoretical models. Both semi-analytic models (SAMs) \cite[e.g.][]{10.1093/mnras/stw2912,10.1093/mnras/stz1810} and hydrodynamical simulations \cite[e.g.][]{2020MNRAS.494.5636W,2022MNRAS.511.5475M,10.1093/mnras/stac1368} have laid the groundwork for such panchromatic studies by exploring galaxy morphologies across both regimes.

In addition to size variations, UV and IR emission can also be spatially offset \citep[e.g.][]{2021MNRAS.503.2622C, Inami, 10.1093/mnras/stab3744, 2024arXiv240207982K, 2026April30}. UV and far-infrared emission provide complementary tracers of star formation, probing unobscured and dust-obscured regions, respectively. By linking the spatial distribution of these tracers across the spectrum, we can study young and evolved stellar populations, dust-obscured star formation, and the thermal reprocessing of radiation. In galaxies with complex dust geometries, these components need not be co-spatial, and offsets between them can provide insight into the structure and evolution of star-forming regions \citep[e.g.][]{ocvirk2024dustuvoffsetshighredshiftgalaxies}.

Furthermore, high-redshift galaxies ($z > 5$) are known to exhibit clumpy and irregular morphologies in the rest-frame UV \cite[e.g.][]{Jiang_2013,10.1093/mnras/stab3744}, which can bias centroid measurements by shifting the inferred UV position away from the centre of potential or the brightest star-forming region. This adds an additional layer of complexity when interpreting UV-FIR spatial offsets.

This has important implications for observational studies, as many SED fitting models assume that energy absorbed in the UV is re-emitted locally in the IR \citep[e.g.][]{bagpipes, 2019A&A...622A.103B}. This assumption breaks down in galaxies with significant spatial offsets between these components \citep[e.g.][]{2015MNRAS.452..235S}. Recently \citet{2024arXiv240207982K} found that nearly 30\% of galaxies show offsets between [CII] emission line, far-infrared continuum, optical, and ultraviolet emission. These offsets are often attributed to dust obscuration of the brightest star-forming regions \citep{2026April30}, where centrally concentrated dust can shift the apparent stellar and UV emission away from the true mass distribution, although the underlying physical drivers remain difficult to constrain observationally. Both AGN activity and stellar feedback are expected to contribute to the spatial separation between young star-forming regions and the dust continuum.

While such observations provide valuable insights into the spatial structure of dust and stellar emission, they remain limited by small sample sizes, selection biases, and the inability to directly probe intrinsic galaxy properties \citep[e.g.][]{2007MNRAS.377..523C,Casey_2014,2023MNRAS.526.2665S,2026A&A...707A..29C,2025arXiv250720190V}. To robustly study these offsets, we require statistically significant galaxy samples drawn from large cosmological volumes, while also achieving sufficient resolution to capture the complex physical processes involved. This combination is naturally enabled by zoom simulation suites such as \flares, which simulate large cosmological volumes at coarse resolution and subsequently re-simulate selected regions at much higher resolution. This approach allows us to resolve the small-scale physics driving spatial offsets while retaining a representative range of galaxy environments.

We utilise this strength of zoom simulations by processing \textsc{Flares} galaxies with \textsc{skirt} radiative transfer code \citep{CAMPS201520}, generating pixel-by-pixel spectral energy distributions from UV to FIR. We use Monte Carlo radiative transfer to accurately map radiation exchange between stars and dust due to attenuation and extinction. The representative population provided by \textsc{Flares} at the Epoch of Reionisation also agrees well with observational constraints at EoR \cite[see][]{10.1093/mnras/staa3360}. Using \textsc{Flares} we are also able to test the environmental effect on galactic processes driving UV-IR spatial offsets. All analyses are performed on simulated NIRCam and ALMA observations created from the radiative transfer output \cite[first shown in][]{Paurush}, allowing us to accurately determine observational spatial offsets.

The rest of this paper is structured as follows. In Section \ref{sec2}, we describe the \flares\ simulations, the galaxy selection criteria and the generation of mock observations. We also discuss a clump extraction method used for subsequent analysis. Section \ref{sec3} presents our results of UV-IR spatial offset and their dependence on observational parameters. In Section \ref{sec4} we explore the impact of dust, AGN feedback, star formation and environment on the measured offsets. Finally, our conclusions are presented in Section \ref{sec:conc}.

Throughout this work, we adopt a Planck year 1 cosmology, with neutrinos neglected. The respective values of the cosmological constants used are $\Omega_m = 0.307$, $\Omega_{\Lambda} = 0.693$, $h$  $=$  $0.6777\ $.

\section{Methodology}
\begin{figure}
\centering
   \includegraphics[width=7.5cm]{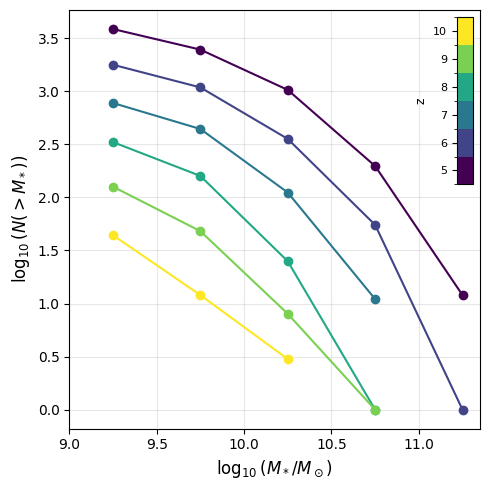}
     \caption{The distribution of galaxies (number counts on y-axis) across different redshifts (plotted as colour bar) and mass bins (x-axis).}
     \label{fig:sample}
     
\end{figure}
\begin{figure*}
   \includegraphics[width=18cm]{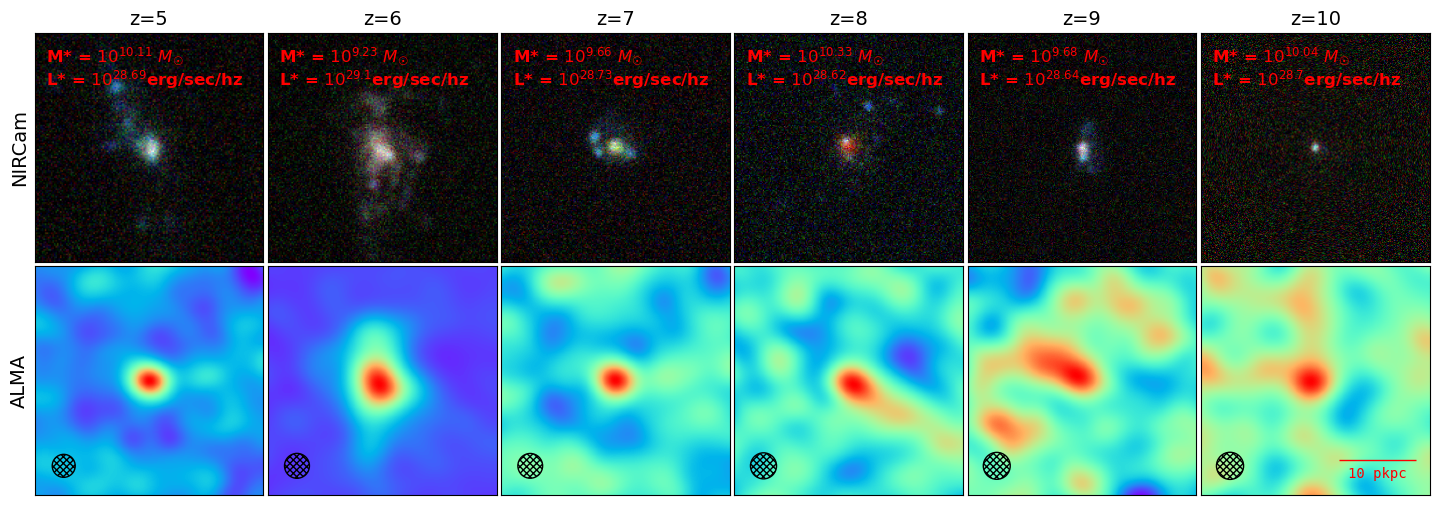}
     \caption{\textbf{Top:} False colour images with red, green and blue channels taken from F200W, F150W, F115W simulated filter data for NIRCam, respectively. \textbf{Bottom:} Simulated ALMA images at rest-frame 158 \um . We label the stellar mass and luminosity for all the samples with the scale shown on the bottom right panel. ALMA beam sizes are shown by the etched circles.}
     \label{fig:false}
     
\end{figure*}
\label{sec2}
\begin{figure*}
   \includegraphics[width=18cm]{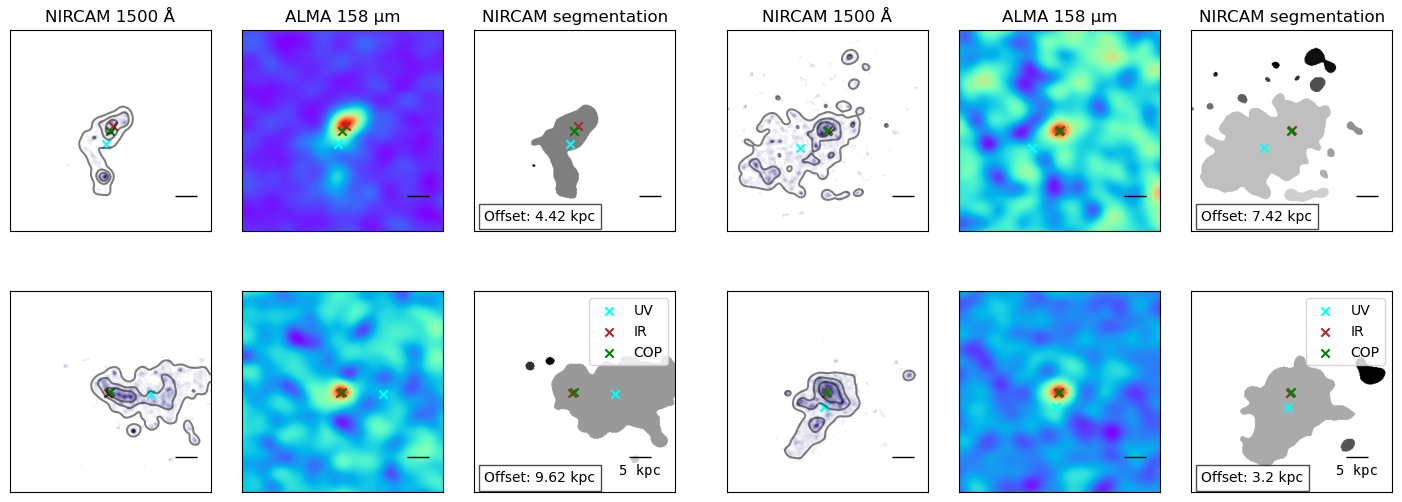}
     \caption{NIRCam and ALMA imaging for four galaxies with significant offset (>2.5kpc). The blue, red and green cross mark the UV, IR centroid fit with \texttt{statmorph} and the centre of potential respectively.}
     \label{fig:example}
     
\end{figure*}

\subsection{First Light And Reionisation Epoch Simulations} \label{flaresintro}
The First Light And Reionisation Epoch Simulations \cite[\flares;][]{10.1093/mnras/staa3360,10.1093/mnras/staa3715} are a suite of zoom simulations of 40 regions chosen from a $3.2$ cGpc on a side dark matter-only box. These regions were re-simulated until $z = 4.67$ with full hydrodynamics using the AGNdT9 configuration of the \eagle\ \citep{10.1093/mnras/stu2058,10.1093/mnras/stv725} galaxy formation model \cite[see Table 3 in][]{10.1093/mnras/stu2058}. This configuration produces similar $z = 0$ stellar mass functions to the reference \eagle\ model but leads to better hot gas properties of groups and clusters compared to the reference \eagle\ model \citep{10.1093/mnras/stx1647}. It also adopts a larger AGN heating temperature increment than the reference model, resulting in fewer but more energetic AGN feedback events and consequently more episodic heating of the surrounding gas. Efficient central black-hole accretion in the \eagle\ model occurs primarily in haloes with masses $\gtrsim 10^{12}\,{\rm M_\odot}$ \citep{10.1093/mnras/staa3360}. Such haloes are absent from the \flares\ population at $z=10$ and constitute less than $0.2\%$ of the population at $z=5$, such that AGN feedback is expected to be important primarily for the rarest and most massive systems at the redshifts considered here.

The regions are selected at $z = 4.7$ and have a radius of 14 ${\rm Mpc}/h$, spanning a wide range of overdensities, from $\delta = -0.479$ to $0.970$. Overdensity ($\delta$) is defined as
\begin{equation}\label{eq:1}
    \delta(\textbf{x}) = \frac{\rho(\textbf{x})}{\bar\rho} - 1,
\end{equation}
where $\rho$ is the average density within the selection sphere at grid coordinates $\textbf{x}$, and $\bar\rho$ is the mean density in the box \cite[see Table A1 in][]{10.1093/mnras/staa3360}.

The simulations were run with a heavily modified version of PGADGET-3, an N-Body Tree-PM smoothed particle hydrodynamics (SPH) code \cite[last described in][]{Springel2005}. The selection of primarily overdense regions helps produce more massive galaxies in the simulation. This selection is not fully representative of the Universe, so a weighting scheme \cite[described in \S 2.4 in][]{10.1093/mnras/staa3360} is used to calculate composite distribution functions that are representative of the total parent volume.

\subsection{Galaxy selection}

 A Friends-of-Friends (FoF; \citealt{1985ApJ...292..371D}) algorithm is first applied to the simulation to identify gravitationally bound groups of particles. Within each FoF group, galaxies in \flares\ are identified using the \textsc{Subfind} algorithm \citep{10.1046/j.1365-8711.2001.04912.x,2009MNRAS.399..497D}, which locates saddle points in the density field of a FoF halo to isolate self-bound substructures. The position of the most bound particle within each substructure is taken to represent the centre of the galaxy.
The stellar mass of each galaxy is then measured from all star particles enclosed within a spherical aperture of radius 30~pkpc centred on this point. To ensure that galaxies are both physically well resolved and suitable for radiative transfer calculations, we follow the same selection criteria as \citet{Vijayan_2022}, selecting only systems containing more than 1000 bound star particles. This requirement corresponds to galaxies with stellar mass, $\text{M}_{\star}\ \gtrsim 10^{9.00}$ \(\textup{M}_\odot\) and UV luminosity $\text{L}_{\text{UV},1500 \AA} > 10^{26.9}$ $\text{erg}$ $\text{s}^{-1}$ $\text{Hz}^{-1}$. The distribution of the selected sample is shown in Figure~\ref{fig:sample}.
Unless stated otherwise, all spatial offset measurements presented in this work are performed at $z = 5$, and at $z > 5$ we used the most massive progenitor identified.
The \flares\ galaxy sample reproduces observed galaxy sizes across multiple wavelengths \citep{10.1093/mnras/stac1368,Paurush} and shows good agreement with observed colour-colour distributions \citep{Vijayan_2022,2022MNRAS.517.3227W}.
\subsection{Mock image creation} \label{Imagecreation}

We described the mock observational pipeline used for NIRCam and ALMA simulations in \citet{Paurush}. Here we give a brief overview of the pipeline. \skirt\ \citep{CAMPS201520,camps2020skirt}, is setup to simulate radiation transfer with a \citet{Weingartner_2001} SMC type dust mixture to account for the low metallicity of galaxies at these redshifts. The effect of self-absorption by dust, as well as CMB heating, is taken into account. The simulations run with $10^6$ photon packets for each source. The observation instruments in \skirt\ are placed at a distance of 1 Mpc from the galaxy, and have a field of view of 60x60 pkpc$^2$ spread over 400x400 pixels. For more details on parameter choices in \skirt\ refer to \citet{Vijayan_2022}. 

Using the per-pixel Spectral Energy Distribution (SED) output from \skirt\ for both UV and IR spectra, we create mock images at FUV (1500 \AA) for NIRCam and IR photometry at 50 \um\ and 250 \um. For these images, the observational effects of PSF and noise are also added. We also simulate ALMA images for dust continuum around 158 \,\um. We use CASA \citep{CASA}, \texttt{simobserve} to create the measurement set for our images. The SEDs of young stellar populations (age $\le10$ Myr) are modelled using the MAPPINGS III \cite[see][for more details]{MappingsIII} template. These were produced using the MAPPINGS III photo-ionisation modelling code to model the nebular emission from young stars. The model assumes the STARBURST99 SPS code \cite[]{Starburst99} for the stellar tracks and a Kroupa IMF \cite[]{Kroupa2002}. 
The older populations (age $>10$ Myr) are modelled using the BPASS \cite[see][for more details]{BPASS2018} SPS code, assuming the Chabrier IMF \citep{chabrier}. Figure \ref{fig:false} shows a sample of simulated NIRCam and ALMA observations for galaxies at redshifts, $z\in[5,10]$.

\subsection{Morphological property calculation}

For the derivation of morphological parameters from our images, we primarily used 
\texttt{statmorph} \citep{10.1093/mnras/sty3345} library. This is an open source code library that measures morphological properties of galaxies from photometry. It provides us with the centre of the object, its radius at various light fractions, concentration, asymmetry, clumpiness and Gini-M20 statistics (see \citealt{Conselice2003} for definitions). In this study, we used \texttt{statmorph} on our mock images to get UV and IR observational centroids. In case of multiple objects detected in imaging, the object with the highest SNR ratio is used for morphological properties.

\subsection{Clump extraction}
We also quantify the clumpiness of the galaxies from the images at this epoch. To extract and count the clumps we use the algorithm defined in \citet{Zanella_2019}. As clumps are anomalies to the symmetric light distribution in a galaxy, we use the S{\'e}rsic profiles generated by \texttt{statmorph} to get the S{\'e}rsic residuals. For clumps to affect the observed centres and morphology, they have to be significantly bright and star-forming, hence we cleaned our S{\'e}rsic residual of data below 5 $\sigma$ detection. We used the \texttt{detect\_sources} class from the \texttt{photutils} \citep{larry_bradley_2023_7946442} library to create a segmentation map, and count objects with at least 5 connected pixels without any deblending.

\section{Results}\label{sec3}

We further investigate how observational effects — in particular, the signal-to-noise ratio (SNR) of \textit{JWST}/NIRCam and ALMA imaging — can drive the observed centres away from the centre of potential (COP, the most gravitationally bound particle as derived by \textsc{Subfind}) in the simulation (Section \ref{obseffect}).  Then we analyse the spatial offsets between the ultraviolet and IR centres of galaxies (Section \ref{UV-IR}).  We examine the correlation between these spatial offsets and key intrinsic galaxy properties to better understand the physical and observational drivers behind the observed trends.

\subsection{Observational effects on UV-IR centroids} \label{obseffect}
\begin{figure}
\centering
   \includegraphics[width=8.5cm]{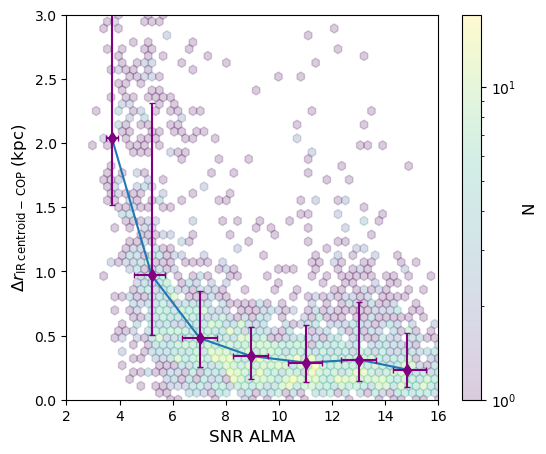}
     \caption{The distance between observational infrared centroid and the centre of potential (y-axis) is plotted against signal to noise ratio of ALMA observations (x-axis).}
     \label{fig:almasnr}
     
\end{figure}

Galaxies at high redshift ($z$ > 5) can have complex and clumpy structures \cite[e.g.][]{Jiang_2013,10.1093/mnras/stab3744} in the rest frame UV which, when resolved at high SNR, draws away the asymmetric fit centroid from the centre of potential and also the brightest pixel. To see if galaxy centres in observations depend on signal-to-noise ratio we compare the observational effects of centroids fit by \texttt{statmorph}. Figure \ref{fig:example} shows that the UV centroid fit by \texttt{statmorph} can lie in low surface brightness regions of the galaxy. At low SNRs, low-luminous brightness regions of the galaxy are not resolved, and the centroid being fit shifts towards brighter emissions. To analyse this effect of SNR we calculated the UV-IR offsets at NIRCam SNRs of 5, 10, 15 and 20. 

\begin{table}
    \centering
    \begin{tabular}{cccc}
    \hline
        \multicolumn{4}{c}{\textbf{$\Delta r_{\rm UV\,\text{centroid} - \,COP}\;(\mathrm{kpc})$}}\\
        \hline
        \textbf{SNR 5}&\textbf{SNR 10} & \textbf{SNR 15}& \textbf{SNR20}\\
        &\textbf{(kpc)} & \textbf{(kpc)}& \textbf{(kpc)}\\\hline \hline
         $0.28^{0.67}_{ 0.12}$& $0.28^{0.82}_{0.11}$& $0.29^{0.95}_{0.10}$& $0.30^{1.02}_{0.10}$\\ \hline
    \end{tabular}
    \caption{Median, the 16th and 84th percentile  offset between UV observational centroid and COP at various NIRCam SNRs}\label{tab:snroffset}
\end{table}

Table \ref{tab:snroffset} shows the median, 16th and 84th percentile UV-COP offsets at these SNR values. The median and 16th percentile offset differ drastically with SNR, while the 84th percentile offset increases with increasing SNR. As low surface brightness regions of the galaxy become resolved, the morphological structure of the galaxy starts affecting the centroid fitting for clumpy galaxies, which increases the number of galaxies with high UV-IR centroid offset, increasing the 84th percentile. 

In the case of interferometric ALMA imaging, as discussed in \citet{Paurush}, at low SNRs, noise around the emission can act as resolved structures and increase the size of the galaxy. Since dust in \flares\ galaxies is found in compact cores \cite[see][]{10.1093/mnras/stac1368,FLARES9,Paurush}, this extended emission can increases asymmetry and offset the IR fit centroid to bright dusty cores. Figure \ref{fig:almasnr} plots the infrared centroid-COP offset against the ALMA SNRs. Due to adopting fixed sky temperatures, zenith opacity and fixed observing duration, we observed with the SNR in the range of [3,15]. The scatter and the median IR-COP offset increase to  2-3 kpc at SNR < 6. At low ALMA SNRs, noise acting as extended emission appends asymmetrical structure to the galaxy, leading to offset IR centroids. This effect reduces at SNR > 8 where the offset medians and the scatter seem invariant with increasing SNR.

To minimise the centre being offset due to clumpy resolved low-surface brightness emission or interferometric noise in all major analyses, we use the brightest pixel as the centre in both UV and IR imaging. A fixed observational plane, instead of face-on or edge-on imaging, used in this study also makes our sample span a wide range of observational angles. It is important to note that at the highest redshifts ($z\ge9$) -- corresponding to the most massive galaxies (M$_{\star}\ge 10^{10}$ M$_{\odot}$), where there is a dearth in the number of galaxies sampled in \flares\  we would see an increased effect of the viewing angle. 

\subsection{UV-IR spatial offsets} \label{UV-IR}

We use imaging generated in Section \ref{Imagecreation}, to calculate UV-IR spatial offset by probing the distance between the two centres. For UV, we use the NIRCam filter corresponding to rest frame 1500 \AA\ and for IR, we use ALMA images at rest frame 158 \um\ at a resolution of $\approx 0.3"$.
As discussed in Section \ref{obseffect}, to minimise centre offset due to asymmetry, the offset between UV and IR emission is defined as the distance between the UV-IR emission peaks. We use the centre of potential, as the true centre of the galaxy to compare our offsets. To avoid outliers in our data, we also remove cases where the UV-IR offset is greater than the sum of half-light radii at the wavelengths probed. Figure \ref{fig:example} shows the imaging and the calculated centre in both UV and IR, along with the centre of potential. All galaxies in the plot show a significant ($>2.5$ kpc) offset\footnote{This definition of significant offset being $>2.5$ kpc is used further in all analyses.}. Table \ref{tab:offset} also presents the mean-median offset statistics for our sample of galaxies. The UV-IR spatial offset is primarily driven by the UV centre being offset from the COP, as the mean UV-COP offsets are significantly higher than the IR-COP offsets. 

\begin{figure}
\centering
   \includegraphics[width=7.5cm]{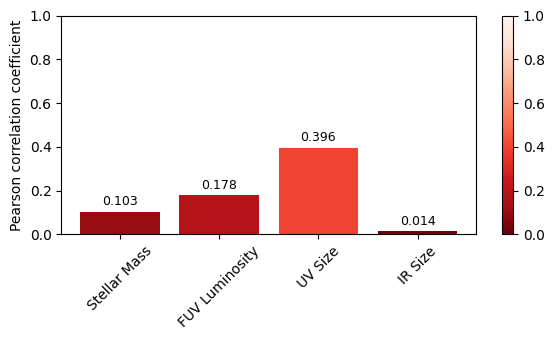}
     \caption{Pearson correlation coefficient between UV-IR offset and galaxy properties.}
     \label{fig:offset_corr}
\end{figure}

Spatial offset also shows no correlation with stellar mass, FUV luminosity or galaxy sizes (UV or FIR) as shown by Figure \ref{fig:offset_corr}, which details the Pearson correlation coefficient of UV-IR spatial offset with galaxy properties.
\begin{table}
    \centering
    \begin{tabular}{cccc}
    \hline
        & \textbf{UV-IR} & \textbf{UV-COP} & \textbf{IR-COP}\\
        & \textbf{Offset (kpc)} & \textbf{Offset (kpc)} & \textbf{Offset (kpc)}\\\hline \hline
Median & 0.2693 & 0.2529 & 0.0537 \\
16th & 0.0921 & 0.0744 & 0.0446 \\
84th & 0.9743 & 0.9200 & 0.1163 \\
Mean & 0.5748 & 0.5442 & 0.1131 \\
Std-Dev & 0.8453 & 0.8125 & 0.2921 \\
    \hline
    \end{tabular}
    \caption{Offset between UV and IR along with their offset with COP at SNR = 20 photometry (most gravitationally bound particle in the simulation).}
    \label{tab:offset}
\end{table}

\begin{figure*}
   \includegraphics[width=17.5cm]{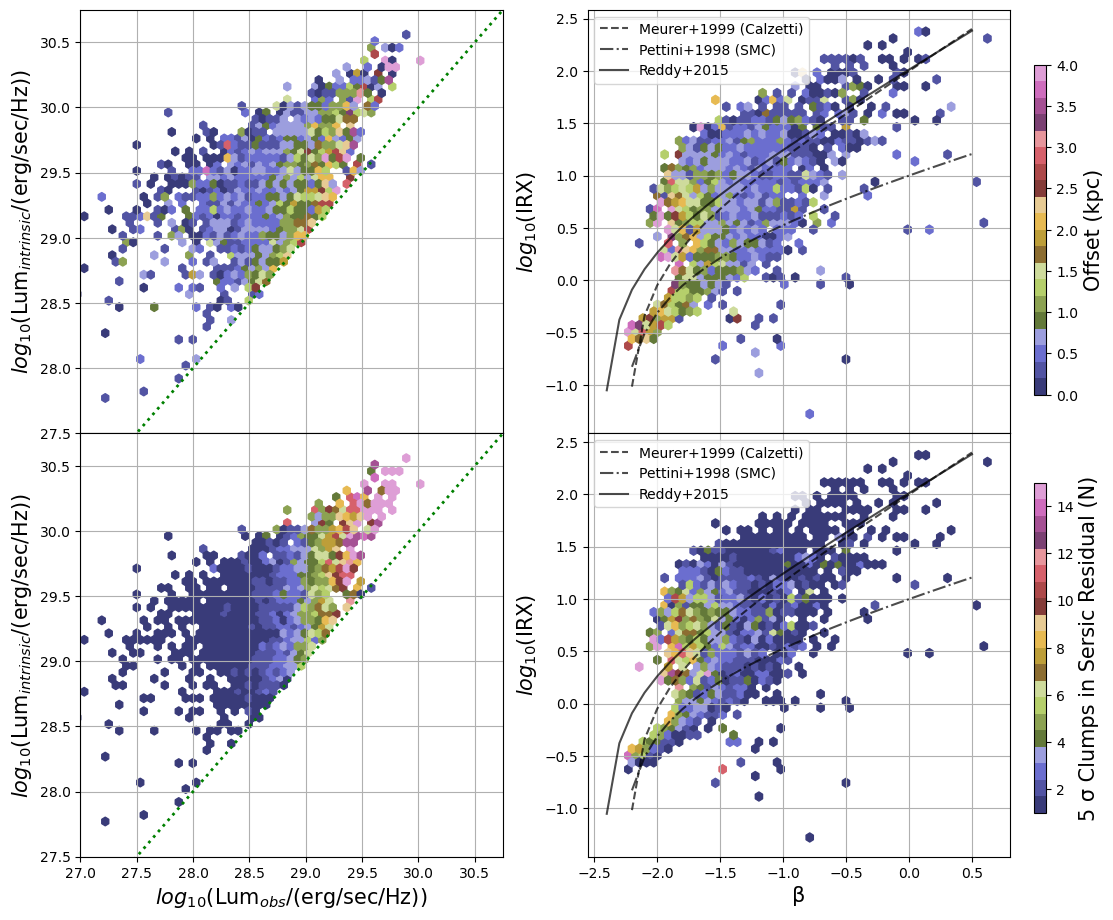}
     \caption{\textbf{Left:} The intrinsic UV (1500 \AA) luminosity (y-axis) is plotted against the dust-attenuated luminosity (x-axis) calculated by radiative transfer. {Right:} The sample is plotted on the IRX-$\beta$ plane (y and x-axis respectively). The top row is coloured by the UV-IR offset, while the bottom row is coloured according to the number of 5 $\sigma$ clumps in UV S{\'e}rsic residual. The green dotted line on the left panes shows the 1:1 relation while the black curves show the dust attenuation curves. Empirical relation for dust screen models for SMC \citep{Pettini} and Calzetti \citep{Calzetti, 1997AJ....114.1834C}, and from \citet{Reddy} are also shown.}
     \label{fig:effectofdust}
     
\end{figure*}
\section{Discussion}\label{sec4}
To analyse the physical mechanisms which drive spatial offset between different wavelengths, we analyse the variation of spatial offset with galaxy evolution mechanisms of dust-radiation interaction, star formation, environment and AGN feedback in this section.
\subsection{Effect of dust}

The distribution of dust in a galaxy can lead to variability in the amount of dust extinction, decreasing the observed UV luminosity for dusty regions compared to the intrinsic emission \citep[e.g.][]{Witstok2023,Shivaei,Markov}. We look at the difference in intrinsic and observed emission for our sample to map the effect of dust attenuation on UV-IR spatial offset. In Figure \ref{fig:effectofdust} (left subplot) we plot the no-dust UV luminosity against the radiative transfer luminosity binned in UV-IR spatial offset bins. We also plotted our galaxies on a plane of IRX-$\beta$ to further analyse the effect of dust attenuation. IRX is the infrared excess defined as the ratio of the total infrared luminosity ($L_{\mathrm{IR}}$) to the UV luminosity ($L_{\mathrm{UV}}$).
\begin{gather}\label{eq: IRX}
	\mathrm{IRX} = \frac{L_{\mathrm{IR}}}{L_{\mathrm{UV}}} \simeq \frac{\int_{8\mumetre}^{1000\mumetre}L_{\lambda}\mathrm{d}\lambda}{L^{'}_{\mathrm{1500}}},
\end{gather} 
where $L^{'}_{\mathrm{1500}}=L_{\mathrm{1500}}\times1500$ \AA, with L$_{\mathrm{1500}}$ being the far-UV luminosity calculated at $1500$ \AA. 
$\beta$ is the UV-continuum slope defined as:
\begin{gather}\label{eq: beta}
	\beta = \frac{\mathrm{log}_{10}(L_{\mathrm{1500}}/L_{\mathrm{2500}})}{\mathrm{log}_{10}(1500/2500)} - 2,
\end{gather}
where $L_{\mathrm{1500}}$ and $L_{\mathrm{2500}}$ are the luminosity of far-UV and near-UV, respectively. 
We notice that most of the high-offset (> 2.5 pkpc) sample of galaxies have high luminosities ($ > 10^{28.5}$ erg/sec/Hz). Importantly, the sample which shows the highest variation in intrinsic and observed luminosities are devoid of high offset galaxies. These galaxies lie close to 1:1 intrinsic and observed UV luminosity ratio with a variation smaller than 1 dex. On the IRX-$\beta$ plane, the high-offset galaxies lie closer to a Calzetti-like attenuation relation \citep{Calzetti,1997AJ....114.1834C} than to an SMC-like curve. Given that our radiative transfer calculations adopt an SMC-type dust grain mixture, this displacement reflects the combined effects of the adopted dust extinction properties and the complex, spatially inhomogeneous star–dust geometry. Spatially offset galaxies span a wide range of IRX but lie on the bluer range of $\beta$ ([-2.5,-1.5])\footnote{UV attenuation and the exact location of individual galaxies on the IRX-beta plane are dust-model dependent. \citet{Vijayan_2022} described how assuming a MW dust mixture would lead to $\beta$ values which are always negative whereas the total IR luminosity and the wavelength of the peak of the IR emission remain in-variate compared with SMC dust mixture.}, indicative of recent starburst or patchy star formation with concentrated and compact dust cores. An off-centre concentration of recent star formation can shift the UV emission away from a centrally concentrated, dust-obscured region, while the FIR emission remains concentrated around the dust-rich centre. \cite{10.1093/mnras/stac1368,FLARES9} have shown abundance of concentrated dust in the galaxy centre is common in massive galaxies due to concentrated localised starbursts.

These starburst clumps can be segmented as separate objects or connected by low surface brightness dispersed regions of the galaxies. As the S{\'e}rsic profile of a galaxy describes how the intensity of light varies at a given distance from the centre of a galaxy, we tried to extract and count the number of segments in the S{\'e}rsic residual of the NIRCam images to quantify the clumpiness of a galaxy. Figure \ref{fig:effectofdust} (right subplot) plots the galaxies binned in count of $5 \sigma$ clumps in the S{\'e}rsic residual for the luminosity profiles and IRX-$\beta$ relation. The clumpy galaxies occupy a similar location the IRX-$\beta$ plane.

\subsection{Effect of star formation}

\begin{figure}
   \includegraphics[width=8.5cm]{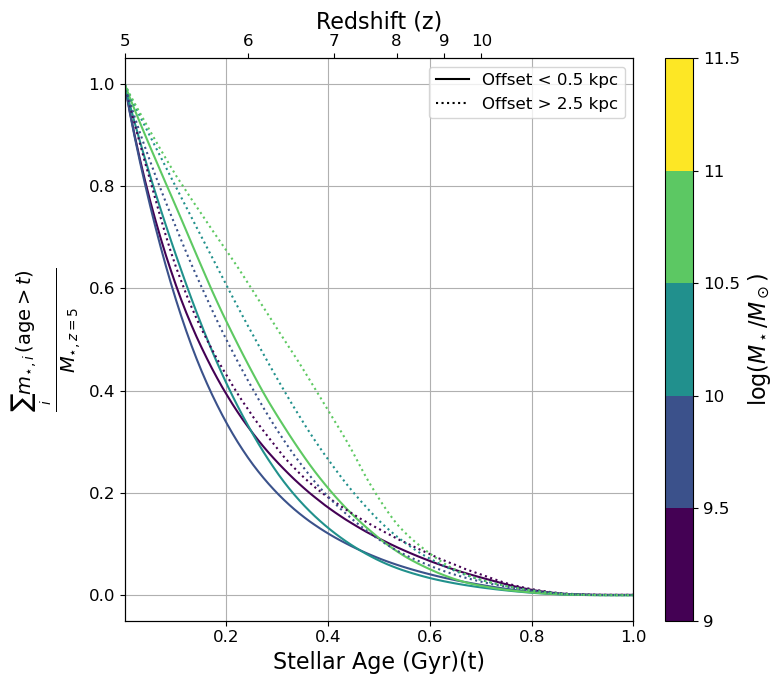}
     \caption{The ratio of cumulative stellar mass (y-axis) of all stellar particles older than stellar age on x-axis to the final galaxy mass at z=5 is plotted in bins of UV-IR offset. Galaxies showing significant offset (> 2.5 kpc) are plotted with dotted lines while the rest of the sample is plotted with solid lines.}
     \label{fig:sfhistory}
     
\end{figure}

To analyse the effect of star formation on UV-FIR spatial offset, we examine the stellar ages of all particles in galaxies at $z=5$. Figure \ref{fig:sfhistory} shows the normalised stellar mass growth histories in bins of stellar mass (measured at $z=5$), along with the age distribution of galaxies exhibiting UV-FIR offsets greater than 2.5 pkpc. Galaxies with larger offsets display enhanced star formation at earlier times ($z>7$), with this trend becoming more pronounced at higher stellar masses. In particular, galaxies with $\mathrm{M}_{z=5}^{\star}\geq 10^{10.5}$ assemble half of their final stellar mass nearly 0.1 Gyr earlier than galaxies with smaller offsets.

As shown in \citep{FLARES9} in the \flares\ framework, such early and sustained star formation is associated with the gradual build-up of metals beyond the central regions of galaxies over cosmic time. This can promote efficient gas cooling at larger radii, facilitating the emergence of inside-out growth in compact systems. In this context, the galaxies exhibiting large UV--FIR offsets are consistent with systems in which star formation has become increasingly spatially extended, enhancing the likelihood that the UV emission is dominated by regions offset from the dust-rich core.

\begin{figure}
   \includegraphics[width=8cm]{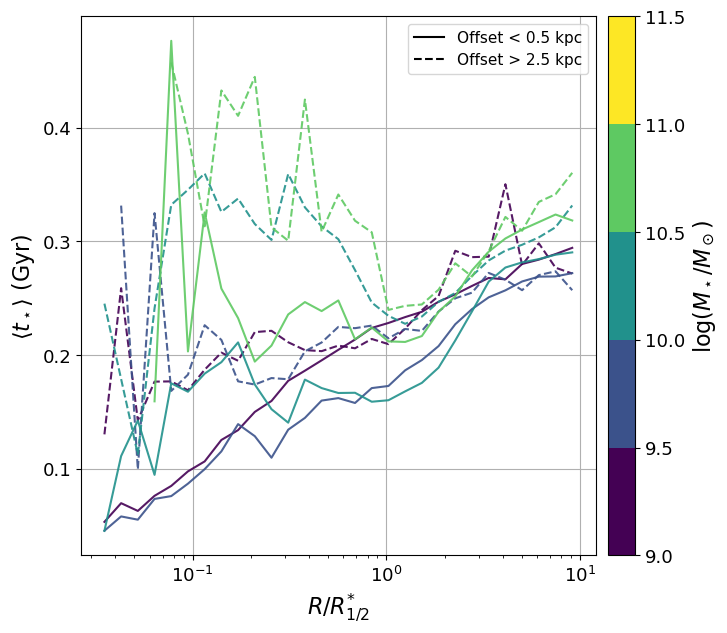}
     \caption{Median stellar ages (y-axis) at a given radius (x-axis, normalised with half-mass radius) is plotted for our sample binned in stellar mass and UV-IR offset. Galaxies showing significant offset (> 2.5 kpc) are plotted with dotted lines while the rest of the sample is plotted with solid lines.}
     \label{fig:radstelage}
     
\end{figure}

More broadly, \citet{FLARES9} show that \flares\ galaxies at $z \approx 5$ exhibit a diversity of evolutionary pathways, including both inside-out growth in compact systems and outside-in evolution in initially diffuse galaxies that later develop centrally concentrated star formation. Such evolutionary transitions can lead to spatial decoupling between young star-forming regions and dust causing the UV-FIR spatial offsets observed in this work. This growth also corresponds well to the bluer UV spectral slope (See Figure \ref{fig:effectofdust}) for galaxies with a significant offset between their UV and IR centres. To understand this transition with respect to UV-IR offset, we also analysed the radial stellar age profiles of these galaxies. Figure \ref{fig:radstelage} plots the median stellar ages of star particles at a given radius from the centre of the potential normalised by the half-mass radius. Galaxies without a significant UV-IR offset have their recent star formation constrained to the cores, while the galaxies with an offset show higher median stellar ages till their half-mass radius. High offset galaxies have an earlier transition to a period of rapid stellar growth.

\begin{figure*}
   \includegraphics[width=17.5cm]{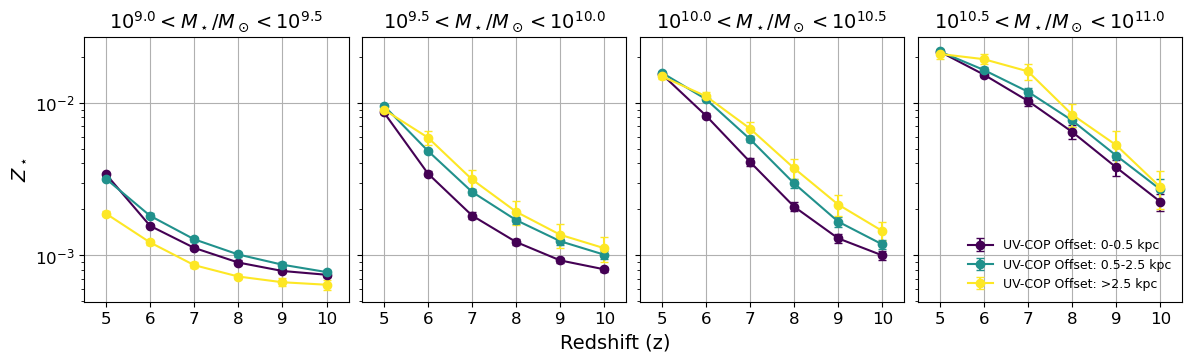}
     \caption{Stellar metallicities (y-axis) evolution with redshift (x-axis) is plotted for z = 5 galaxies. The evolution is tracked by considering the most massive progenitors at z>5. The sample is binned in stellar mass (shown in different panes) and in UV-COP offset (shown by different colours).}
     \label{fig:metalevol}
     
\end{figure*}

This is further validated when we analyse the metallicity evolution by looking at the progenitors of our sample. Figure \ref{fig:metalevol} plots the stellar metallicities against redshift for our sample at $z=5$, and its progenitors binned in UV-IR offset bins. For galaxies that show an inside-out growth ($\text{M}^{\star}\geq 10^{9.5}$), higher offset bins have higher metallicities at each redshift. Fast and early enrichment of the ISM with metals leads to an early transition to inside-out growth. 

AGN feedback can also drive gas and dust away from star-forming cores \citep{feedback1,feedback2,feedback3}. In principle, this can suppress star formation in the central regions while allowing enriched gas at larger radii to cool and form stars, thereby shifting the UV emission away from the centre of potential. However, this mechanism is not expected to be prominent for the majority of galaxies in our sample. The AGNdT9 configuration adopted by \flares\ produces fewer but more energetic AGN feedback events, with efficient central black-hole accretion occurring primarily in haloes with masses $\gtrsim 10^{12}\,{\rm M_\odot}$. Such haloes are absent from the \flares\ population at $z=10$ and constitute less than $0.2\%$ of the population at $z=5$ \citep{10.1093/mnras/staa3360}. Furthermore, as shown in Appendix \ref{AGN feedback}, we find no systematic difference in black-hole growth or accretion between galaxies with and without significant UV--IR offsets. We therefore do not find evidence that AGN-driven gas removal or suppression of central star formation is the dominant mechanism responsible for the offsets in our sample.

\subsection{Effect of environment and mergers}
\begin{figure}
   \includegraphics[width=7.5 cm]{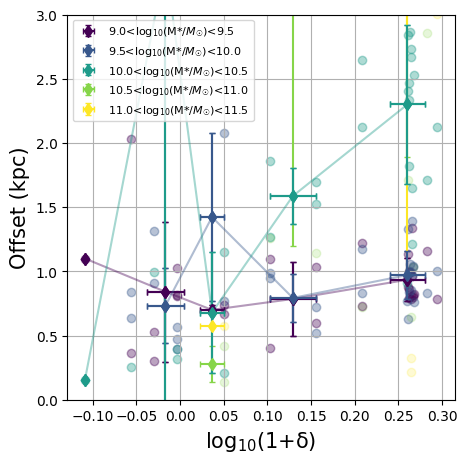}
     \caption{UV offset (y-axis) is plotted against $\log{(1+\delta)}$ where $\delta$ is the overdensity (x-axis) of the \flares\ region. The faint circle plots show the average offset for each of the 40 regions in each stellar mass bin. The solid diamond plots show the median when the data is binned in $\log{(1+\delta)}$.}
     \label{fig:enviro}
     
\end{figure}
Galaxy interactions and mergers can lead to disk instabilities and localised star formation away from the gravitational centre of the galaxies. The possibility of these interactions changes with the density of the environment. \flares\ has regions of different densities, and so we analysed if UV-IR offset is affected by the environment. In Figure \ref{fig:enviro}, we binned our galaxies in 0.5 dex stellar mass bins and plotted the evolution of UV-IR offset against $\log_{10}(1+\delta)$ where $\delta$ is the overdensity (defined in equation~\ref{eq:1}).
We see no significant evolution of median UV-IR offset with increasing density of environment. 

Major mergers in galaxies can affect the distribution of gas and dust in a galaxy. This redistribution of gas can trigger localised starbursts outside the present stellar distribution. These newly formed stars would remain unobscured, while some pre-merger stellar distribution can be heavily dust attenuated. Such a variation can result in an observed spatial offset between UV and IR. Any observation during or close to mergers will also contain multiple cores of galaxies, which can lead to a case where the dust centre and the UV centre are selected for different merging galaxies. 

For our sample at $z=5$, using merger graphs built using \mega\ \citep{MEGA}, if any galaxy had two progenitor sub-halos which contributed more than a quarter of the stellar mass at the next snapshot, we define a major stellar merger to have happened. The galaxies with very high offset have an undistinguishable merger history compared to the rest of the sample. The galaxies with offset less than 2.5 pkpc, had a stellar major merger fraction of 18.82$\%\ $ while the galaxies with higher offsets also had a major merger fraction of 18.61 $\%\ $. While major stellar mergers do not correlate with the presence of UV-IR offset, it could be possible that a galaxy accumulated a lot of gas through minor mergers with large gas-to-stellar mass ratios. This accumulated gas, when cooled, could lead to star formation outside the dense cores. If any galaxy had two progenitor sub-halos which contributed more than half the gas mass at the next snapshot, we define a major gas merger as having taken place.  If any galaxy had two progenitor sub-halos which contributed more than half the gas mass at the next snapshot, we define a major gas merger as having taken place. In our case, such a major gas merger also seems invariable with offset. The galaxies with offset less than 2.5 pkpc had a major gas merger fraction of 11.60$ \%\ $ while the galaxies with higher offsets also had a major merger fraction of 11.99 $\%$.

Halo mass plays a key role in shaping the distribution of gas and dust, as well as the star formation activity of galaxies. In more massive haloes, dust becomes increasingly centrally concentrated, resulting in stronger attenuation in the inner regions. This enhanced obscuration can lead to a displacement of the observed UV centroid from the core. \citet{ocvirk2024dustuvoffsetshighredshiftgalaxies}
analyses the UV-IR offset using the COSMIC DAWN III simulations \citep{2022MNRAS.516.3389L} and finds that the mean simulated offsets increase with dark matter
halo mass at $z$ = 5. While our median offsets remain in agreement with \citet{ocvirk2024dustuvoffsetshighredshiftgalaxies} (Figure \ref{fig:hmass} plots the halo mass against the UV-IR offset), we notice an increase in median offsets with increasing halo mass at $\text{M}_{\text{halo}}\geq 10^{11} \textup{M}_\odot$). Although due to high scatter, we are not able to confidently predict any evolution.

\section{Conclusions}\label{sec:conc}

In this study, we used the simulated observations created in \citet{Paurush} by post-processing \flares\ galaxies with using \skirt\ \citep{CAMPS201520,camps2020skirt} Monte Carlo radiative transfer code. We used the simulated observations to study UV-FIR spatial offset. We looked at the observational factors that affect the offset. We also studied the effect of feedback and galaxy growth mechanisms, which contribute to these offsets.
Our main findings from these studies are as follows:
\begin{enumerate}
    \item About 16\% of galaxies show an offset greater than $\geq 2.5$ kpc between their UV and IR observed UV and FIR centroids. The centres fit by \texttt{statmorph} depend on the morphology of the galaxy. When low surface brightness clumpy or dispersed structures are resolved, asymmetry increases, leading to higher offsets. The brightest pixel of the emission can be used as a better alternative to centroid fitting. Steeper inclination angles for observations also reduce the offset. 

    \item Galaxies which show a significant offset (>2.5 kpc) lie on the bluer end of the UV spectral slope ($-2.5<\beta<-1.5$), indicating recent star formation. These galaxies also lie along the Calzetti attenuation curve \citep{Calzetti} showing patchy star-dust geometry causes this IRX-$\beta$ behaviour rather than the intrinsic dust properties. Dust-attenuated cores lead to UV extinction, drawing away the centre to more dispersed star-forming regions.

    \item Galaxies with a significant offset exhibit a faster evolution and an early transition to an inside-out or outside-in growth. These galaxies form 50\% of their final $z$ = 5 stellar mass nearly 0.1 Gyr (at $z \sim 7$) earlier than galaxies that don't show an offset. These galaxies also have 2-3 times higher radial stellar ages below their half mass radius compared to their counterparts. This faster evolution and early transition are fuelled by faster enrichment of the ISM, with these galaxies having higher stellar metallicities throughout their evolution.

    \item UV-IR spatial offset shows no correlation with AGN feedback or density of the environment. Galaxies with higher offset also show a merger history statistically similar to the rest of the population. 
\end{enumerate}

These results provide a physical interpretation for the spatial offsets observed in high-redshift galaxies. Observational studies have reported extended dust emission and spatial offsets between UV and FIR tracers, often interpreted as evidence for distinct star-forming regions or complex dust geometries. Our analysis suggests that such offsets can arise naturally from a combination of dust-obscured central star formation and clumpy, spatially distributed UV emission, without requiring a direct correspondence between UV and FIR emitting regions. The preference for bluer UV slopes in high-offset systems indicates that these galaxies are undergoing recent, spatially extended star formation, while centrally concentrated dust obscuration shifts the observed UV centroid away from the core. 

Furthermore, the lack of strong correlations with halo mass, AGN activity, or environment implies that these offsets are primarily governed by internal processes such as star formation history, metal enrichment, and dust geometry, rather than large-scale structure. This is consistent with recent observational findings that highlight the role of patchy dust distributions and wavelength-dependent morphologies in driving apparent UV-FIR size differences and offsets. Overall, \flares\ demonstrates that the observed diversity in UV-FIR spatial structure can be explained by the interplay between star formation and dust within galaxies, and that caution must be exercised when interpreting UV and IR emission as co-spatial tracers of star formation.

\section*{Acknowledgements}
We thank the anonymous referee for
their helpful comments which helped make this manuscript better.

This work has been performed using the Danish National Life Science Supercomputing Centre, Computerome. This work also used the DiRAC@Durham facility managed by the Institute for Computational Cosmology on behalf of the STFC DiRAC HPC Facility (www.dirac.ac.uk). The equipment was funded by BEIS capital funding via STFC capital grants ST/K00042X/1, ST/P002293/1, ST/R002371/1 and ST/S002502/1, Durham University and STFC operations grant ST/R000832/1. DiRAC is part of the National e-Infrastructure. This work also made use of the University of Hertfordshire's high-performance computing facility (uhhpc.herts.ac.uk). The Cosmic Dawn centre (DAWN) is funded by the Danish National Research Foundation under grant DNRF140. APV, WJR, and SMW acknowledge support from the Sussex Astronomy Centre STFC Consolidated Grant (ST/X001040/1). 
      
We also wish to acknowledge the following open source software packages used in the analysis: NUMPY \citep{Harris_2020}, SCIPY \citep{Virtanen_2020}, PANDAS \citep{reback2020pandas}, ASTROPY \citep{astropy:2013,astropy:2018,astropy:2022}, PILLOW \citep{clark2015pillow}, MATPLOTLIB  \citep{4160265} and STATMORPH \citep{10.1093/mnras/sty3345}.

 We list here the roles and contributions of the authors according to the Contributor Roles Taxonomy (CRediT) \footnote{\href{https://credit.niso.org/}{https://credit.niso.org/}}. \textbf{Paurush Punyasheel:} Conceptualization, Data Curation, Formal analysis, Investigation, Methodology, Visualization, Writing – original draft. \textbf{Aswin P. Vijayan:} Conceptualization, Data Curation, Methodology, Supervision, Writing – review \& editing. \textbf{William J. Roper:} Conceptualization, Data Curation, Methodology, Writing – review \& editing. \textbf{Thomas R. Greve:} Supervision, Writing – review \& editing. \textbf{Hiddo Algera, Steven Gillman, Bitten Gullberg:} Methodology, Writing – review \& editing. \textbf{Christopher C. Lovell:} Data Curation, Methodology, Writing – review \& editing. \textbf{Shihong Liao, Robert M. Yates, Stephen M. Wilkins:} Writing – review \& editing.

\section*{Data Availability}

The morphological analysis data is available at  \href{https://github.com/paurush-p/FLARES_XXII_plots}{https://github.com/paurush-p/FLARES\_XXII\_plots}. Raw data from the simulations can be made available upon request.



\bibliographystyle{mnras}
\bibliography{example} 



\FloatBarrier
\appendix

\section{Halo mass relation with spatial offset}

\begin{figure}
\centering
   \includegraphics[width=7.5 cm]{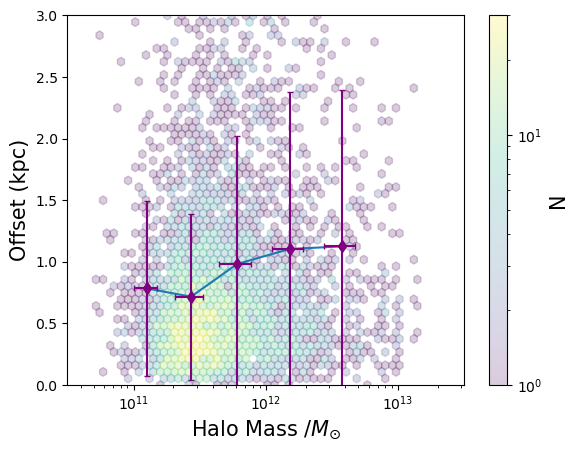}
     \caption{UV-FIR spatial offset(y-axis) versus halo mass (x-axis) at $z=5$. The hexbin map shows the density of galaxies, while overplotted points indicate the mean offset in halo mass bins with standard deviation. Despite a weak increase at high halo mass, the large scatter suggests no significant correlation.}
     \label{fig:hmass}
     
\end{figure}

Figure \ref{fig:hmass} shows the relationship between halo mass and the UV-FIR spatial offset for our galaxy sample. While the full distribution spans a wide range of offsets at all halo masses, there is no clear correlation between halo mass and offset. The scatter remains large across the entire halo mass range, indicating that halo mass alone is not a strong predictor of the UV-FIR spatial offset.

We also compute the median offset in bins of halo mass and find only a weak increase towards higher halo masses ($\mathrm{M}_{\mathrm{halo}} \gtrsim 10^{11}\,\mathrm{M}_\odot$). However, this trend is not statistically significant given the large dispersion in each bin. Overall, our results suggest that the spatial offset is primarily driven by internal galaxy processes rather than global halo properties. 

\section{Effect of AGN feedback} \label{AGN feedback}

\begin{figure}
\centering
   \includegraphics[width=7.5cm]{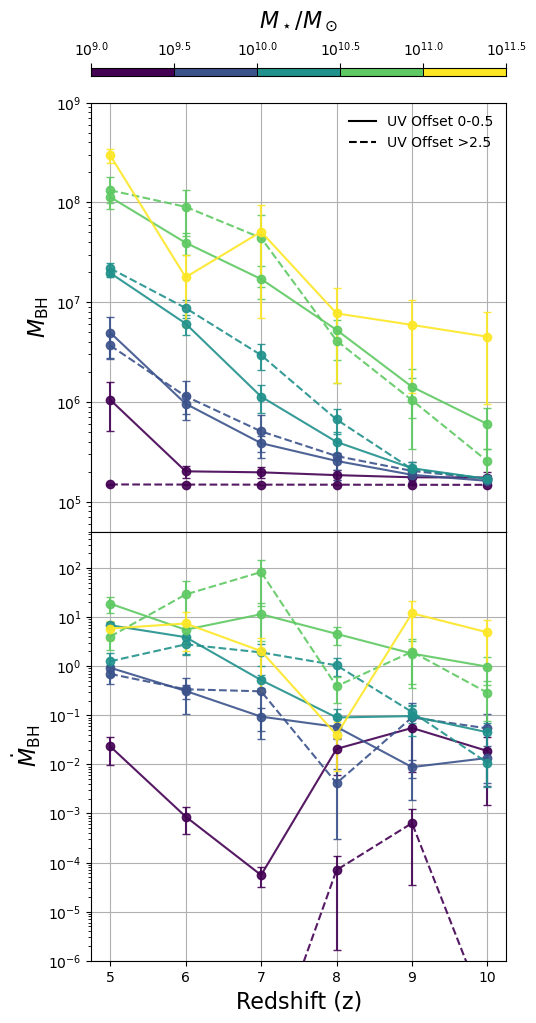}
     \caption{The evolution of black hole mass ($M_{\mathrm{BH}}$, top panel) and accretion rate ($\dot{M}_{\mathrm{BH}}$, bottom panel) as a function of redshift ($z = 5$--$10$). Tracks are color-coded by galaxy stellar mass bins ($M_\star/M_\odot$), while line styles distinguish no offset systems ($\text{Offset } 0\text{--}0.5\text{ kpc}$, solid) and highly offset systems ($\text{Offset } >2.5\text{ kpc}$, dashed).}
     \label{fig:feedback}
\end{figure}

Figure \ref{fig:feedback} plots the evolution of blackhole mass and accretion rate for galaxies with an offset by comparing it to the blackhole mass and accretion rates of the 0-0.5 kpc offset bin. There are no clear trends with redshift for either of the features. This confirms that AGN feedback does not play a major role in creating the UV-FIR offset.

\bsp	
\label{lastpage}
\end{document}